\documentclass[lettersize,journal]{IEEEtran}
\usepackage{amsmath,amsfonts}
\usepackage{dsfont}
\usepackage{array}
\usepackage{textcomp}
\usepackage{stfloats}
\usepackage{url}
\usepackage{verbatim}
\usepackage{graphicx}
\usepackage{cite}

\DeclareMathAlphabet\mathchorus     {T1}{qzc} {m} {n}
\usepackage[linesnumbered, ruled]{algorithm2e}
\usepackage{amsmath}
\usepackage{multirow}
\usepackage{multicol}
\usepackage{tikz}
\usepackage{pgfplots}
\usepackage{amssymb}
\usepackage{subcaption}
\usepackage{hyperref}
\usepackage{amsthm}
{}
{}
{}
\usetikzlibrary{patterns}
\usepgfplotslibrary{statistics}
\pgfplotsset{compat=1.8}
\pgfmathdeclarefunction{fpumod}{2}{%
    \pgfmathfloatdivide{#1}{#2}%
    \pgfmathfloatint{\pgfmathresult}%
    \pgfmathfloatmultiply{\pgfmathresult}{#2}%
    \pgfmathfloatsubtract{#1}{\pgfmathresult}%
    \pgfmathfloatifapproxequalrel{\pgfmathresult}{#2}{\def\pgfmathresult{5}}{}%
}
\pgfplotsset{boxplot legend/.style={
    legend image code/.code={
        \draw[#1] (0cm,-0.1cm) rectangle (0.4cm,0.1cm)
        (0.2cm,-0.1cm) -- (0.2cm,-0.2cm) (0.05cm,-0.2cm) -- (0.35cm,-0.2cm)
        (0.2cm,0.1cm) -- (0.2cm,0.2cm) (0.05cm,0.2cm) -- (0.35cm,0.2cm);
     \path (0cm,0.24cm) (0cm,-0.24cm);  
    },
}}
\makeatletter
\newcommand{\removelatexerror}{\let\@latex@error\@gobble}
\def\ps@IEEEtitlepagestyle{%
	\def\@oddfoot{\mycopyrightnotice}%
	\def\@oddhead{\hbox{}\@IEEEheaderstyle\leftmark\hfil\thepage}\relax
	\def\@evenhead{\@IEEEheaderstyle\thepage\hfil\leftmark\hbox{}}\relax
	\def\@evenfoot{}%
}
\def\mycopyrightnotice{%
	\begin{minipage}{\textwidth}
		\centering \scriptsize
		This article has been accepted in IEEE Wireless Communication Letters Journal © 2023 IEEE. Personal use of this material is permitted. Permission from 
		IEEE must be obtained for all other uses, in any current or future media, including reprinting/republishing this material for advertising or promotional purposes, creating new collective works, for resale or redistribution to servers or lists, or reuse of any copyrighted component of this work in other works. This work is freely available for survey and citation.
		
	\end{minipage}
}
\makeatother

\begin{document}

\title{An AI-driven Intelligent Traffic Management Model for 6G Cloud Radio Access Networks}

\author{Smruti Rekha Swain, Deepika Saxena, Jatinder Kumar,	Ashutosh~Kumar~Singh,~\IEEEmembership{Senior~Member~IEEE},~and~Chung-Nan~Lee,~\textit{Member, IEEE}
	\thanks{ S. R. Swain, J. Kumar, and A. K. Singh are with the Department of Computer Applications, National Institute of Technology, Kurukshetra, India. E-mail: smruti.sai90@gmail.com, jatinderkumar2851@gmail.com,  ashutosh@nitkkr.ac.in \\
	 D. Saxena is with Department of Computer Science, Goethe University Frankfurt, Germany. E-mail: 13deepikasaxena@gmail.com\\
	  C. N. Lee is with the Department of Computer Science and Engineering, National Sun Yat-Sen University, Kaohsiung, Taiwan. E-mail: cnlee@cse.nsysu.edu.tw)} 
}

\markboth{IEEE Wireless Communications Letters}%
{Shell \MakeLowercase{\textit{et al.}}: A Sample Article Using IEEEtran.cls for IEEE Journals}


\maketitle

\begin{abstract}
This letter proposes a novel Cloud Radio Access Network (C-RAN) traffic analysis and management model that estimates  probable RAN traffic congestion and mitigate its effect by adopting a suitable handling mechanism. A computation approach is introduced to classify heterogeneous  RAN traffic into distinct traffic states based on bandwidth consumption and execution time of various job requests. Further,  a cloud-based traffic management is employed to schedule  and allocate resources among user job requests according to the associated traffic states to minimize latency and maximize bandwidth utilization.   The experimental evaluation and comparison of the proposed model with state-of-the-art methods reveal that it is effective in minimizing the worse effect of traffic congestion and improves bandwidth utilization and reduces job execution latency  up to 17.07\% and 18\%, respectively.
\end{abstract}

\begin{IEEEkeywords}
C-RAN, Congestion, Stragglers, Network Hogs.
\end{IEEEkeywords}
\vspace{-.80cm}
\section{Introduction}
\IEEEPARstart{B}{ooming} C-RAN traffic (i.e., RAN traffic based on a centralized cloud
computing used in the deployment of 5G networks to provide improved cellular coverage, capacity, and reliability) has become inevitable with the emerging technologies, including 6G wireless communication networks, Internet of Things (IoT), and unmanned aerial vehicle (UAV) assisted networks, etc \cite{zhao2020federated}.  According to a recent report, the C-RAN market size is projected to grow from USD 5.5 billion in 2022 to USD 43.2 billion by 2032, at a Compound Annual Growth Rate (CAGR) of 22.9\% during the forecast period \footnote{https://www.factmr.com/report/cloud-radio-access-network-market}. C-RAN allows  virtualization of networks and customize them to cater the specific needs of applications, services, devices, and customers  \cite{kim2019task}. However, the diversity and complexity of  6G C-RAN environment impose a critical and challenging issue of  wireless traffic management because of conflicting perspectives of customers and service provider where, the former aspires for highest service quality with least latency and the latter intends minimize operational cost by maximizing bandwidth utilization.  In this context, a novel {\textbf{T}raffic \textbf{M}anagement model for 6G \textbf{C}loud \textbf{R}adio} \textbf{A}ccess \textbf{N}etwork  (\textbf{TM-CRN}) is proposed that examines the heterogeneous RAN traffic  for probable network congestion and filters the contention-suspected traffic from the normal traffic to facilitate appropriate management.  Firstly, the RAN traffic is estimated and analysed proactively for classification of live job requests into distinct traffic states. Thereafter, cloud-based traffic management is employed which triggers the allocation of job requests confined to various traffic states according to their criticality and resource requirement such that job execution latency is minimized and resource (viz., CPU, memory, and bandwidth) is maximized. \textit{The key contributions are}: (i) A RAN traffic analyser is introduced that estimates the respective traffic proactively to  detect and mitigate the probable congestion while reserving sufficient capacity of physical resources.
 (ii) A novel approach of RAN traffic states classification, is determined for resource allocation, scheduling, and mapping accordingly.
 (iii) TM-CRN model is implemented and the results are verified
against  QoS metrics, and compared with state-of-the-art approaches.

\vspace{-.250cm}
\section{TM-CRN Model} 
\begin{figure}[!htbp]
    \centering
    \includegraphics[width=1.05\linewidth,scale=2]{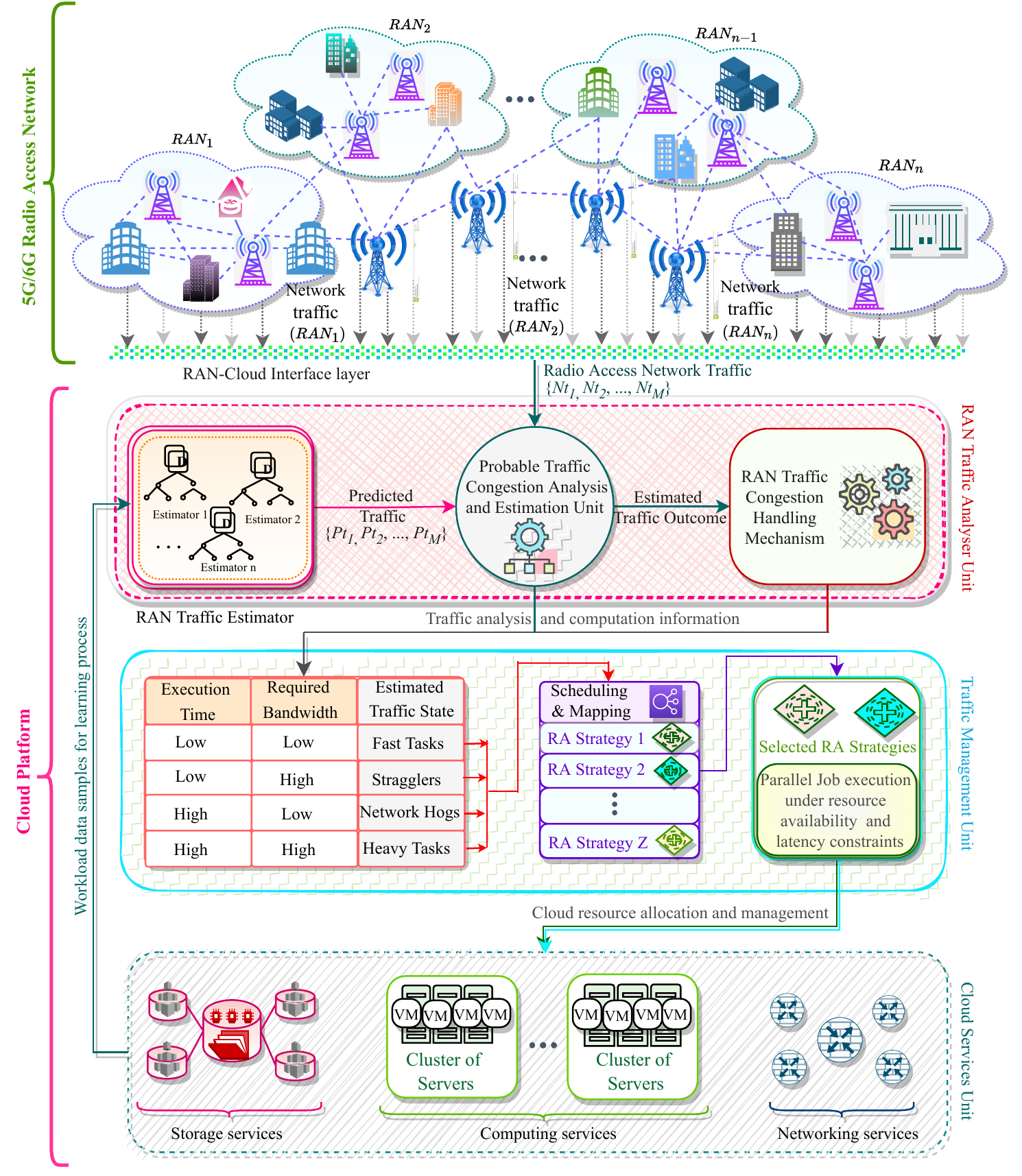}
    \caption{TM-CRN Architectural Design}
    \label{fig:my_labelpm}
\end{figure}

Consider a  wireless telecommunications environment such as Open-RAN, GSM-RAN or C-RAN \cite{habibi2019comprehensive} composed of antennas, radios, and baseband units (BBU)s:  \{$RAN_1$, $RAN_2$, ..., $RAN_n$\}$\in \mathds{RAN}$ producing heterogeneous network traffic \{$Nt_1$, $Nt_2$, ..., $Nt_M$\}$\in \mathds{NT}$ as illustrated in  Fig. \ref{fig:my_labelpm}. In C-RAN, BBU devices convert digital signals into radio transmissions and vice-versa, are acting as a centralized control and processing station which are  connected to remotely located Radio Frequency Units (RFU, the broadcasting antenna of a base station) via high speed optical fiber.
The traffic includes job requests varying over execution time and resource demands such as bandwidth ($BW$), CPU ($C$), and memory ($Mem$), are transferred to cloud platform or BBU for different purposes including computation, storage, data sharing, etc., via \textit{RAN-Cloud Interface Layer}.  
The cloud platform  consists of three co-operative operational units, including \textit{RAN Traffic Analyser Unit} (RTAU), \textit{Traffic Management Unit} (TMU), and  \textit{Cloud Services Unit} (CSU). The radio network traffic ($\mathds{NT}$) is passed to RTAU for probable traffic congestion identification and estimation using the knowledge generated by AI-driven \textit{RAN Traffic Predictor} i.e., \{$Pt_1$, $Pt_2$, ..., $Pt_M$\} and pre-estimated threshold of traffic deviation. Accordingly, the estimated traffic outcome is transferred to \textit{RAN Traffic Congestion Handling Mechanism} which triggers the essential operations to manage it as depicted in TMU block. Within TMU, $\mathds{NT}$ is segregated into distinct traffic states such as \textit{Fast jobs}, \textit{Stragglers},  \textit{Network Hogs}, and \textit{Heavy jobs}, etc., depending up on the amount of  execution time and bandwidth consumption. The most appropriate single or multiple scheduling and mapping strategies for the live job requests are decided among the $Z$ resource allocation strategies, and applied as per the fulfilment of essential resource availability and latency criteria. Further, TMU explores the selected resource allocation strategies for parallel execution to minimize the resource wastage, execution time and reduce CPU idle time. Correspondingly, the job requests are executed by CSU comprising of \textit{Storage}, \textit{Computing}, and \textit{Networking} services. The cloud service execution information like, resource ($C$, $BW$, $Mem$) usage, number and type  of job requests per time-interval is utilized to generate training data samples for the learning process of RAN traffic predictor.  The operational description of RTAU and TMU with CSU is given in the Section \ref{rtau} and Section \ref{ctmu}, respectively. 
\vspace{-.25cm}
\section{RAN Traffic Analysis} \label{rtau}
As illustrated in Fig. \ref{fig:my_labelpm}, RTAU analyses approaching traffic using RAN traffic predictor (RTP) which is based on the AI-driven {Extreme-Gradient Boosting} (XGB) algorithm. This predictor is capable of  learning and developing the intuitive and precise correlations among extracted samples or patterns. Let the RTP  is composed of $\mathchorus{l}$ base learners i.e., decision trees $ \mathchorus{BL}^\ast=\{\mathchorus{BL}^\ast_1, \mathchorus{BL}^\ast_2, ..., \mathchorus{BL}^\ast_l\}$ which estimate $\mathchorus{l}$ respective outcomes  $\mathchorus{O}^\ast=\{\mathchorus{O}^\ast_1, \mathchorus{O}^\ast_2, ..., \mathchorus{O}^\ast_l\}$  using Eq. (\ref{eq:xgb1});
\begin{equation}
  \label{eq:xgb1}
	\mathchorus{O}^\ast=\sum_{z=1}^{l}\mathchorus{BL}_z^\ast(\mathchorus{F}_i) \quad \forall i \in \{1, 2, ..., s\}  
\end{equation}
where $\mathchorus{F}$ represents the input vector of size $s$, representing $s$ training  attributes (such as bandwidth, memory, processing usage samples) for traffic prediction.   During each iteration, decision trees are trained incrementally to reduce prediction errors and the amount of error reduction  is computed as \textit{gain} or  \textit{loss term} ($L(\mathchorus{O}^\ast, {\mathchorus{O}^{\ast\ast}_{t-1}} + {\mathchorus{BL}^\ast_t}(\mathchorus{F}_i))$). Concurrently, a \textit{regularisation term} ($\Psi(\mathchorus{BL}^\ast_t)$) is computed at each iteration to measure the complexity of decision trees and control their pruning in optimized way while preventing overfitting. Accordingly,  the objective function ${L}$ composed of loss term and regularisation term  is minimized using  Eq. (\ref{eq:xgb2}). \vspace{-.25cm}
\begin{equation}
\label{eq:xgb2}
	{L}_t = \sum_{i=1}^{s} {L(\mathchorus{O}^\ast, {\mathchorus{O}^{\ast\ast}_{t-1}} + {\mathchorus{BL}^\ast_t}(\mathchorus{F}_i)) + \sum_{z=1}^{l}\Psi(\mathchorus{BL_z}^\ast)}   
\end{equation}
The term $\Psi(\mathchorus{BL}^\ast_t)$ is calculated using Eq. (\ref{eq:xgb5}), where $\gamma$ and $\lambda$ are $L_1$ and $L_2$ regularisation coefficients, respectively, $w$ is internal split tree weight and  $K$ is number of leaves in tree.
\begin{equation}
 \label{eq:xgb5}
 	\Psi(\mathchorus{BL}^\ast_t)=\gamma K + \dfrac{1}{2} \lambda ||w||^2
\end{equation}
Thereafter, adaptive boosting or optimization is applied using Taylor expansion which calculates exact loss for each decision tree by transforming Eq. (\ref{eq:xgb2}) into Eq. (\ref{eq:xgb3});
\vspace{-.25cm}
\begin{gather}
  \label{eq:xgb3}
  {L}_t = \sum_{i=1}^{ s^\ast} { [g_i\mathchorus{BL}^\ast_t(\mathchorus{F}_i) + \frac{1}{2}h_i {{\mathchorus{BL}^\ast_t}}^2(\mathchorus{F}_i)] + \Psi(\mathchorus{BL}^\ast_t)}  
\end{gather}  
where $g_i=\partial_{\mathchorus{O}^{\ast\ast}_{t-1}}{L(\mathchorus{O}^\ast, {\mathchorus{O}^{\ast\ast}_{t-1}})}$, and $h_i=\partial^2_{\mathchorus{O}^{\ast\ast}_{t-1}}{L(\mathchorus{O}^\ast, {\mathchorus{O}^{\ast\ast}_{t-1}})}$ are the first and second order derivatives of loss function in the gradient, respectively (more details about XGBoost can be found in \cite{chen2016xgboost}).  During each time-interval $\{t_a, t_b\}\in t$, $a<b$, live traffic values are provided as input to RTP for traffic prediction in the next time interval  $\{t_{a+1}, t_{b+1}\}\in t$, $a<b$. 

Let $\mathchorus{Tr}_\sigma$ be the deviation of network traffic from the predicted traffic ($Pt$) over duration $\Delta{t} \in$ \{$t_a$, $t_b$\}, $\mathchorus{Tr}_\sigma^{thr}$ and $\Delta{t}^{thr}$ are traffic deviation and time-period thresholds respectively. Eq. (\ref{uba}) investigates traffic status ($\Xi^{status}$) for duration  ($\Delta{t}$) based on the comparison between expected traffic deviation ($\mathchorus{Tr}_\sigma$) and threshold ($\mathchorus{Tr}_\sigma^{thr}$). 
\begin{equation}
  \label{uba}
\Xi^{status} \times \Delta{t}  = \begin{cases}
 1, & {\text{If}(\mathchorus{Tr}_\sigma \times \Delta{t}  > \mathchorus{Tr}_\sigma^{thr} \times \Delta{t}^{thr})} \\
 -1, & {\text{If}(\mathchorus{Tr}_\sigma \times \Delta{t} < 0)} \\
 0, & {\text{Otherwise}}  
 \end{cases}	
 \end{equation}
Accordingly, the traffic status $\Xi^{status}$ value `1' indicates `\textit{network congestion}', `-1' specifies `\textit{sub-normal traffic}', and `\textit{normal traffic}', otherwise. Further, the critical and sufficient condition for the congestion is formulated and defined in Eq. (\ref{e1}) subject to four constraints ($C_1$-$C_4$)  over time ($\Delta{t}$);
\begin{equation}
 \centering
 \label{e1}
  \begin{aligned}
      \int_{t_a}^{t_b} { \bigg(\sum_{i=1}^{M}{Pt_i}   +  \sum_{i=1}^{M}{{\mathchorus{Tr}_\sigma}_i} } \bigg) dt \leq  \int_{t_a}^{t_b}{\sum_{i=1}^{M}\big({Nt_i} }\big) dt   \\ s.t. \quad \forall{C_1} \bigvee
\exists\{C_2, C_3, C_4\} 
 \\
\left.
\begin{array}{ll}
{C_1:} \quad t_b - t_a = \Delta{t} > \mathchorus{Tr}_\sigma \times \Delta{t}^{thr})\\
C_2: \quad {\sum_{i=1}^{M}{Nt_i^{BW}}} \geq \sum_{k=1}^{P}{PM_k^{BW}} \\
C_3: \quad {\sum_{i=1}^{M}{Nt_i^{C}}} \geq \sum_{k=1}^{P}{PM_k^{C}}\\
C_4: \quad  {\sum_{i=1}^{M}{Nt_i^{Mem}}} \geq \sum_{k=1}^{P}{PM_k^{Mem}}
\end{array}
\right\}  
\end{aligned}
\end{equation}
where  $Nt_i^{BW}$, $Nt_i^{C}$, and $Nt_i^{Mem}$ specify capacity demand of  bandwidth, CPU, and memory, respectively, of the live traffic ($Nt$). If the aggregated demand of the entire traffic ($\sum_{i=1}^{M}{Nt_i}$) is greater than estimated resource requirement ($\sum_{i=1}^{M}{Pt_i}   +  \sum_{i=1}^{M}{{\mathchorus{Tr}_\sigma}_i}$), then congestion is detected. $C_1$ states traffic overflow ($\mathchorus{Tr}_\sigma$) for longer than threshold time-period ($ \Delta{t}^{thr}$); $C_2$- $C_4$ state excess demand of resources viz., $BW$, $C$, $Mem$ of the live traffic than the available aggregated capacity of resources on $P$ physical machines ($PM$). If \textit{congestion is anticipated}, the network traffic is diverted across multiple pathways and assigned to physical machines reserved for handling the network hogs and heavy workloads. Likewise, if the approaching traffic is sub-normal ($\Xi^{status}$ = $-1$), the respective job requests are executed on least number of physical machines subject to resource availability constraints; otherwise, the traffic is  \textit{normal} which is handled  by TMU. The traffic handling and management by TMU  is  discussed in the following Section \ref{ctmu}.
\vspace{-0.50cm}
\section{Cloud-based Traffic Management} \label{ctmu}
The RAN traffic analysis and computation information is transferred to TMU to allow estimation of the distinguished traffic states, including \textit{Light jobs} ($\mathchorus{Tr}^{Lit}$), \textit{Stragglers} ($\mathchorus{Tr}^{Stg}$), \textit{Network Hogs} ($\mathchorus{Tr}^{Hog}$),  \textit{Heavy jobs} ($\mathchorus{Tr}^{Hvy}$), and \textit{Average jobs} ($\mathchorus{Tr}^{Avg}$) etc.
Eq. (\ref{states}) determines various states of live RAN traffic \{$Nt_1$, $Nt_2$, ..., $Nt_M$\} at $t^{th}$ instance by considering different circumstances of the  bandwidth demand ($(Nt_i^{BW}: i \in \{1, M\}$)  and  execution time ($Nt_i^{Et}$).         
\vspace{-.250cm}
\begin{multline}\label{states}
\resizebox{0.45\textwidth}{!}{$  
{Nt_i^{State}}=	\begin{cases}
 \mathchorus{Tr}^{Lit} (1), & {\text{If}(Nt_i^{BW}<BW_i^{thr} \quad\&\&\quad Nt_i^{Et} < Et_i^{thr})} \\
 {\mathchorus{Tr}^{Stg}} (2), & {\text{If}(Nt_i^{BW}\leq BW_i^{thr} \quad\&\&\quad Nt_i^{Et} \geq Et_i^{thr})} \\
 {\mathchorus{Tr}^{Hog}} (3), & {\text{If}(Nt_i^{BW}\geq BW_i^{thr} \quad\&\&\quad Nt_i^{Et} \leq Et_i^{thr})} \\
 \mathchorus{Tr}^{Hvy} (4), & {\text{If}(Nt_i^{BW}\geq BW_i^{thr} \quad\&\&\quad Nt_i^{Et} \geq Et_i^{thr})} \\
\mathchorus{Tr}^{Avg} (5) & {\text{Otherwise}}   
\end{cases}$}
\vspace{-.50cm}
\end{multline}
TMU schedules job requests ($Jr$) belonging to distinct traffic states exclusively in the most admissible way to minimize latency of execution and maximize the resource ($BW$, $C$, $Mem$) utilization as stated in Eqs. (\ref{s1}-\ref{s5}) subject to constraints \{$C_1$-$C_6$\} specified in Eq. (\ref{cs}). The expression $\omega_{kji}$ represents mapping among $k^{th}$ job ($Jr_k$: {$k\in M$}), $j^{th}$ virtual node ($VN_j$: {$j\in Q$}), and $i^{th}$ physical machine ($PM_i$: {$i\in P$}); $R$ and $R^{\ast}$ are resource capacity of virtual node and physical machine, respectively. The constraint $C_1$ specifies $k^{th}$ job can be assigned to only one $VN_j$ hosted on one $PM_i$ at an instance;  \{$C_2$-$C_4$\} state resource capacity of $VN_j$ must be lesser or equal to available resource capacity of $PM_i$; and $C_5$ \& $C_6$ specify  resource requirement of $k^{th}$ job ($Jr_k$) for processing must be satisfied by the available resources on the respective $VN_j$ hosted on $PM_i$. The job requests confined to light traffic state ($\mathchorus{Tr}^{Lit}$) are allocated to virtual nodes ($VN$) hosted on  physical machines with required resource capacity using first-come first-serve (FCFS) scheduling (Eq. \ref{s1}). 
\vspace{-.250cm}
\begin{gather}
 {\omega_{kji}^{\mathchorus{Tr}^{Lit}}}=
 Jr_k^{\mathchorus{Tr}^{Lit}} \times VN_j^{R} \times List_{(FCFS)}{PM_i^{R^{\ast}} } \label{s1} \\
 {\omega_{kji}^{\mathchorus{Tr}^{Stg}}}=
 Jr_k^{\mathchorus{Tr}^{Stg}} \times VN_j^{R} \times MAX(List{PM_i^{R^{\ast}}}) \label{s2} \\
 {\omega_{kji}^{\mathchorus{Tr}^{Hog}}}=Jr_k^{\mathchorus{Tr}^{Hog}} \times VN_j^{R} \times MAX(List_
 {BW}{PM_i^{R^{\ast}}}) \label{s3}\\
 {\omega_{kji}^{\mathchorus{Tr}^{Hvy}}}= Jr_k^{\mathchorus{Tr}^{Hvy}} \times \sum_{j=1}^{Z}VN_j^{R} \times \sum_{i=1}^{Z^{\ast}}{PM_i^{R^{\ast}}} \label{s4} \\
  {\omega_{kji}^{\mathchorus{Tr}^{Avg}}}= Jr_k^{\mathchorus{Tr}^{Avg}} \times VN_j^{R} \times MAX(List_{C}{PM_i^{R^{\ast}}}) \label{s5}
\end{gather}

\begin{equation}\label{cs}
\resizebox{0.49\textwidth}{!}{$ 
\centering
\begin{aligned}
\left.
\begin{array}{ll}
\text{subject to} \{C_1-C_6\} \\
{C_1:} \quad \forall_{k \in M}\forall_{j \in Q}\forall_{i \in P}{\omega_{kji}}=1\\
C_2: \quad \forall_{k \in M} \forall_{j \in Q}\forall_{i \in P}{VN_j^{C}} \times \omega_{kji} \leq PM_i^{C^{\ast}}\\
C_3: \quad \forall_{k \in M}\forall_{j \in Q}\forall_{i \in P}{VN_j^{M}} \times \omega_{kji} \leq PM_i^{M^{\ast}}\\
C_4: \quad \forall_{k \in BW}\forall_{j \in Q}\forall_{i \in P}{VN_j^{BW}} \times \omega_{kji} \leq PM_i^{BW^{\ast}}\\
C_5: \quad  \sum_{k\in M}{R}_k  \leq \sum_{i \in P}PM_i^{\mathds{R^{\ast}}} \quad R^{\ast} \in \{C^{\ast}, M^{\ast}, BW^{\ast}\} \\
C_6: \quad  r_k \times {R}_k \leq VN_j^{R^\ast} \quad \forall_k \in [1, M], j \in [1, Q] 
\end{array}
\right\}  
\end{aligned} $}
\end{equation}
Likewise, the stragglers ($\mathchorus{Tr}^{Stg}$)
requiring higher computational capacity  are assigned to virtual nodes hosted on a
physical machine with larger  CPU, bandwidth,
and memory capacity to allow needed I/O operations (Eq. \ref{s2}).
Further, a hybrid scheduling is introduced to allow parallel
execution of stragglers with network hogs ($\mathchorus{Tr}^{Hog}$) such that
the bandwidth hogs can be processed by the computing and
network devices having sufficient resource capacity that can
serve the requirement during idle time when the stragglers
are performing I/O operations (Eq. \ref{s3}). The ‘Heavy jobs’ ($\mathchorus{Tr}^{Hvy}$)
demanding resource capacity larger than the threshold are
bound to execute on multiple physical machines with
highest resource capacity which can altogether  serve resource demand
of the respective traffic (Eq. \ref{s4}). All the remaining job
requests are considered as `average' and `delay-sensitive' to
be scheduled on the remaining physical nodes to allow faster
execution. Specifically, the average jobs ($\mathchorus{Tr}^{Avg}$) based requests are
sorted as per the basis of their deadlines and priorities, and
the physical nodes are assorted in decreasing order of their
processing speed in the list (Eq. \ref{s5}). Such an allocation of virtual nodes
executes lowest deadline requests on highest processing speed
servers to minimize the time of execution.
\section{Operational Design and Complexity}
Algorithm \ref{algo-osecc} imparts the operational summary of TM-CRN for effective traffic management for  cloud-based RAN.
\begin{figure}[!htbp]
	\removelatexerror
	\begin{algorithm}[H]
		\caption{TM-CRN: Operational Summary}
		\label{algo-osecc}
	 \textbf{Input} Number of: PMs ($P$), VNs ($Q$), RANs ($M$) \;
		Initialize: $List_{{Jr}}$, $List_{{VN}}$, $List_{{PM}}$\; 	
		Distribute initial RAN traffic: \{$Nt_1$, $Nt_2$, ..., $Nt_M$\} on VNs hosted at $P$ PMs\;
		\For {each time-interval $\{t_a, t_b\}$}{ 
	Estimate traffic load: $Pt_{t+1}$= RAN Traffic Predictor(${NT}_{t_a}$)\;
	Receive heterogeneous RAN traffic (${NT}_{t_a +1}$) and compute traffic deviation ($\mathchorus{Tr}_{\sigma}$) \;
	Analyse the status of the traffic ($\Xi$) using Eq. (\ref{uba}) \;
	Investigate the exact occurrence of a RAN traffic congestion by applying Eq. (\ref{e1}) \;
	Determine and distinguish among different traffic states using Eq. (\ref{states}) \;
	Apply Eqs. (\ref{s1}-\ref{s5})  to decide allocation of job requests in the most appropriate manner\;
	}
	\end{algorithm}
\end{figure}

Step 1 allows to read the number of PMs, VNs, and RANs' job requests from users, consumes $\mathcal{O}(1)$ complexity while step 2 initializes lists of jobs, VNs, and PMs, has $\mathcal{O}(1)$ complexity. Step 3 schedules $M$ network requests onto $P$ PMs, with $\mathcal{O}(P \times M)$ complexity. Assuming steps 4-11 repeat for $t$ time intervals, wherein step 5 calls RAN traffic predictor   T $\mapsto$ $\mathcal{O}(thzlogn)$ where
$t$ is the number of trees, $h$ is the height of the trees, 
$z$ is the number of non-missing entries in the training, and $n$ is the number of examples. The prediction for a new sample consumes time $\mathcal{O}(th)$.
 Steps 6-10 compute Eqs. (\ref{states}-\ref{s5}) consume $\mathcal{O}(1)$ complexity.  Hence, the total complexity comes out to be $\mathcal{O}(PMT)$.
 \section{Performance Evaluation and Discussion}
 \subsection{Experimental Set-up and Dataset}
The simulation experiments are executed on a server machine assembled with two Intel\textsuperscript{\textregistered} Xeon\textsuperscript{\textregistered} Silver 4114 CPU with 40 core processor and 2.20 GHz clock speed, deployed with 64-bit Ubuntu 16.04 LTS having main memory of 128 GB in Python 3.1. The CDC environment is set up with CoS \cite{saxena2022fault} having IBM servers' CPU (MIPS), RAM (GB) and bandwidth (bps) configurations: \{1060, 2, 2000\}; \{2660, 4, 2000\}; \{3067, 8, 4000\}; \{4076, 64, 8000\}. Four types of VMs inspired from Amazon VM instances with CPU, RAM, and bandwidth: \{500, 1,500\}; \{1000, 2, 1000\}; \{2000, 3, 1000\}; \{2500, 4, 2000\}  are used. We experimented with a wireless traffic data sets of call detail records \cite{barlacchi2015multi} and a C-RAN traffic gathered from Youtube dataset on Mobile streaming  \cite{lohYoutube2022}  containing records of 80 RAN scenarios having  171   bandwidth settings, measured in 1,939 runs with emulated 3G/4G traces.  

\par TM-CRN  is compared with \textit{Federated  Meta-Learning Approach} (FMLA) based wireless traffic prediction \cite{zhang2022efficient}, \textit{Joint UE and Fog Optimization} (JUFO) scheme \cite{kim2019task}, and \textit{Online Secure Communication Model Cloud} (OSC-MC) \cite{saxena2021osc} for different performance metrics. A wireless traffic prediction approach, FMLA is proposed in \cite{zhang2022efficient}  to manage the highly dynamic and low latency wireless communication networks by  learning a sensitive global model with knowledge  gathered from diverse regions. A minimized energy computation derived tasks offloading and C-RAN traffic management scheme is presented in \cite{kim2019task}, wherein cloud tasks priority is investigated to allow the task execution  within the determined delay bound. OSC-MC model \cite{saxena2021osc}  prevents network traffic congestion  and minimize occurrence of network hogs by computing the traffic deviation proactively while maintaining secure workload execution.

\subsection{Numerical Results}
Table \ref{table:2} reports resultant values obtained for the key performance metrics for TM-CRN with varying bandwidth ($BW^{thr}$) and execution-time ($Et^{thr}$) threshold values (including 25\%, 50\%, 75\%, and 90\% of the respective quantitative values)   for 500 C-RAN jobs over progressive time-period. By applying Eq. (\ref{states}), traffic is filtered into different traffic states such that for $BW^{thr}$=25\% and $\mathchorus{Tr}^{Lit}$ (\%) and  $\mathchorus{Tr}^{Hog}$(\%) increase, while $\mathchorus{Tr}^{Stg}$(\%)  and $\mathchorus{Tr}^{Hvy}$(\%) decrease with increasing execution time consumption ($Et^{thr}$) and the similar trend is followed for rest of the $BW^{thr}$ values. Furthermore, with growing $BW^{thr}$, the traffic states $\mathchorus{Tr}^{Lit}$ (\%) and  $\mathchorus{Tr}^{Hvy}$(\%) show reduction, and $\mathchorus{Tr}^{Hog}$(\%) and $\mathchorus{Tr}^{Stg}$(\%) upgrade.  The bandwidth utilization ($BW^{Util}$ \%) is achieved in the range [16\% - 30\%] and  the jobs' execution time ($Et$) varies with their size and  processing speed of cloud servers. However, the significant reduction in overall job execution latency ($\mathchorus{Lt}^{Redc}$) is observed for TM-CRN as compared to the default traffic execution (using first-come first-serve (FCFS) scheduling). The reason is that  TM-CRN efficiently exploits the possible parallelism between jobs confined to $\mathchorus{Tr}^{Hog}$ and  $\mathchorus{Tr}^{Stg}$. 
\begin{table}[!htbp] 
\centering
\caption{Key Performance Indicators for TM-CRN }
\label{table:2}
\resizebox{0.49\textwidth}{!}{
\begin{tabular}{llccccccc}
\hline
\textbf{$BW^{thr}$} & $Et^{thr}$ & $\mathchorus{Tr}^{Lit}$& $\mathchorus{Tr}^{Stg}$ & $\mathchorus{Tr}^{Hog}$ & $\mathchorus{Tr}^{Hvy}$& $BW^{Util}$& $Et^{Cons}\times 10^4$ & $\mathchorus{Lt}^{Redc}$ \\
(\%) & (\%) & (\%)& (\%) & (\%) & (\%)& (\%)  & (sec) &(\%) \\
 \hline
 25&25&11.4  & 24.12& 11.41& 53.07& 19.51 & 18.72&23.80\\
 25&50&41.67 & 05.70& 29.82& 22.81& 17.85 & 22.02&10.43\\
 25&75&56.58  & 01.32& 34.21& 07.90& 20.75 & 23.06&6.17\\
 25&90&64.48  & 0.00& 35.52& 0.00 & 22.46 & 24.58&0.00\\ \hline
 50&25&7.89  & 44.29& 14.91& 32.89& 25.03 & 14.22&42.10\\
 50&50&24.12 & 11.84&47.37& 16.67& 26.16 & 20.06&18.37\\
 50&75&33.77  & 02.19& 57.02& 07.02& 26.11 & 22.41&8.80\\
 50&90&33.77  & 01.75& 57.46& 07.01& 26.11 & 22.59&8.08\\ \hline
 75&25&5.70  & 57.46& 17.10& 19.74& 28.54 & 09.30&62.12\\
 75&50&14.03  & 17.54& 57.46& 10.97& 27.82 & 16.47&32.97\\
 75&75  & 20.61& 04.83&70.18 & 04.38 & 29.07 & 19.48&20.73\\
 75&90&25.00  & 0.00& 75.00& 0.00& 28.82 & 24.58&-0.02\\ \hline
 90&25&01.75 & 67.11& 21.05& 10.08& 18.26 & 06.02&75.52\\
 90&50&06.14 & 22.80& 65.35& 05.70& 25.89& 13.90&43.42\\
 90&75&09.21  & 06.58& 81.57& 02.63& 24.64 & 17.73&27.84\\
 90&90  & 09.21& 06.14& 82.01& 02.63& 24.57&17.91& 27.12
\\ \hline
\noalign{\smallskip}
\end{tabular}}

\end{table}
 
\begin{figure}[!htbp]
    \centering
\begin{tikzpicture}
\begin{axis}[
 width=.38\textwidth,
 height=0.24\textwidth,
    ymin=0,
    ymax=0.6,
    symbolic x coords={ $Mi^{MAE}$, $Mi^{MSE}$, $Tr^{MAE}$,  $Tr^{MSE}$},
    xtick=data,
    ybar,
    ymajorgrids=true,
    xmajorgrids=true,
    grid style=dashdotted,
    bar width=3.5pt,
    enlarge x limits=0.16,
    legend style={at={(1.23,0.98)},
    anchor=north,legend columns=1},
     xlabel={{\textit{Datasets}}},
    ylabel={ \textit{Error}}]
   \addplot+[ybar,fill=orange!20!,thick, mark options={fill=orange}, pattern color=orange!70!, draw=orange!80!, thick,postaction={
        pattern=crosshatch, thin}, thin] plot coordinates{ ({$Mi^{MAE}$}, 0.0015) ({$Mi^{MSE}$},0.001 ) ({$Tr^{MAE}$}, 0.0015) ({$Tr^{MSE}$}, 0.0015)}; \addlegendentry{\small{Optimal }}
   
    \addplot+[ybar,fill=cyan!10!,thin, mark options={fill=white},pattern color=blue!40!, draw=cyan,thick, postaction={
        pattern=crosshatch, thin}, thin] plot coordinates{ ({$Mi^{MAE}$}, 0.0703) ({$Mi^{MSE}$},0.00170 ) ({$Tr^{MAE}$}, 0.2544) ({$Tr^{MSE}$}, 0.04815) }; \addlegendentry{\small{TM-CRN}} 
 \addplot+[ybar,fill=black!10!,thin, mark options={fill=black!10!}, postaction={
        pattern=crosshatch, thin},pattern color=black!50!, draw=black!60!, thin] plot coordinates{ ({$Mi^{MAE}$}, 0.0803) ({$Mi^{MSE}$},0.0170 ) ({$Tr^{MAE}$}, 0.3544) ({$Tr^{MSE}$}, 0.4815) }; \addlegendentry{\small{FMLA-s \cite{zhang2022efficient}}}         

 \addplot+[ybar,fill=blue!6!,thin, mark options={fill=black!10!}, postaction={
        pattern=crosshatch, thin},pattern color=blue!50!, draw=blue!70!, thin] plot coordinates{ ({$Mi^{MAE}$}, 0.0790) ({$Mi^{MSE}$},0.0170 ) ({$Tr^{MAE}$}, 0.3544) ({$Tr^{MSE}$}, 0.4815) }; \addlegendentry{\small{FMLA-w }}     
   \addplot+[ybar,fill=teal!20!,thin, mark options={fill=cyan}, pattern color=teal!70!, draw=teal!80!, thick,postaction={
        pattern=crosshatch, thin}, thin] plot coordinates{ ({$Mi^{MAE}$}, 0.0782) ({$Mi^{MSE}$},0.0169 ) ({$Tr^{MAE}$}, 0.3855) ({$Tr^{MSE}$}, 0.5258)}; \addlegendentry{\small{FMLA-d }}
 
\end{axis}
\end{tikzpicture}
    \caption{Traffic estimation error}
    \label{fig:my_label}
\end{figure}
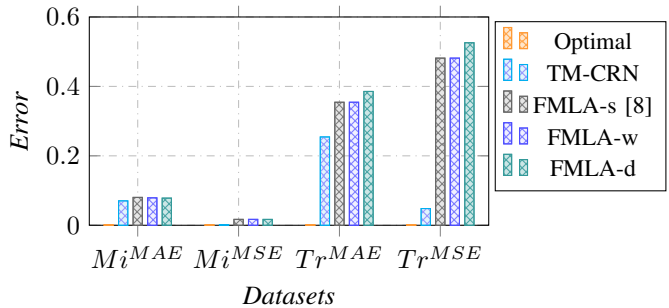

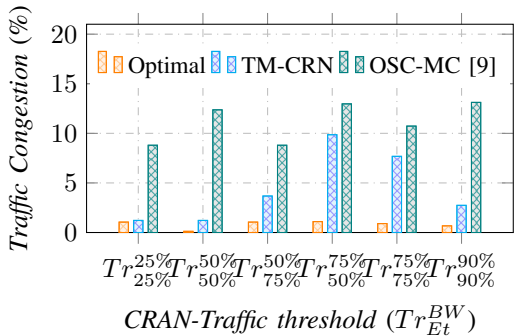
\begin{figure}[!htbp]
    \centering
    
    \begin{tikzpicture}
\begin{axis}[
 width=.40\textwidth,
 height=0.24\textwidth,
    ymin=0,
    ymax=21,
    symbolic x coords={ $Tr^{{25\%}}_{{25\%}}$, $Tr^{{50\%}}_{{50\%}}$, $Tr^{{50\%}}_{{75\%}}$,  $Tr^{{75\%}}_{{50\%}}$, $Tr^{{75\%}}_{{75\%}}$, $Tr^{{90\%}}_{{90\%}}$},
    xtick=data,
    ybar,
    ymajorgrids=true,
    xmajorgrids=true,
    grid style=dashdotted,
    bar width=3.5pt,
    enlarge x limits=0.16,
    legend style={at={(0.5,0.9)},
    anchor=north,legend columns=3, draw=none},
     xlabel={{\textit{CRAN-Traffic threshold} ($Tr^{BW}_{Et}$)}},
    ylabel={ \textit{Traffic Congestion} (\%)}]
    \addplot+[ybar,fill=orange!10!,thick, mark options={fill=white},pattern color=orange!40!, draw=orange,thick, postaction={
        pattern=crosshatch, thin}, thin] plot coordinates{ ($Tr^{{25\%}}_{{25\%}}$, 1.05) ({$Tr^{{50\%}}_{{50\%}}$}, 0.12) ({$Tr^{{50\%}}_{{75\%}}$}, 1.05) ({$Tr^{{75\%}}_{{50\%}}$}, 1.1) ({$Tr^{{75\%}}_{{75\%}}$}, 0.9) ({$Tr^{{90\%}}_{{90\%}}$}, 0.671) }; \addlegendentry{\small{Optimal}}

   \addplot+[ybar,fill=cyan!10!,thin, mark options={fill=white},pattern color=blue!40!, draw=cyan,thick, postaction={
        pattern=crosshatch, thin}, thin] plot coordinates{ ($Tr^{{25\%}}_{{25\%}}$, 1.22) ({$Tr^{{50\%}}_{{50\%}}$}, 1.22) ({$Tr^{{50\%}}_{{75\%}}$}, 3.68) ({$Tr^{{75\%}}_{{50\%}}$}, 9.86) ({$Tr^{{75\%}}_{{75\%}}$}, 7.68) ({$Tr^{{90\%}}_{{90\%}}$}, 2.74) }; \addlegendentry{\small{TM-CRN}}       
   
   \addplot+[ybar,fill=black!10!,thin, mark options={fill=black!10!}, postaction={
        pattern=crosshatch, thin},pattern color=teal!70!, draw=teal!100!, thin] plot coordinates{ ($Tr^{{25\%}}_{{25\%}}$, 8.8) ({$Tr^{{50\%}}_{{50\%}}$}, 12.37) ({$Tr^{{50\%}}_{{75\%}}$}, 8.80) ({$Tr^{{75\%}}_{{50\%}}$}, 12.97) ({$Tr^{{75\%}}_{{75\%}}$}, 10.73) ({$Tr^{{90\%}}_{{90\%}}$}, 13.12)}; \addlegendentry{\small{OSC-MC \cite{saxena2021osc}}} 

\end{axis}
\end{tikzpicture}
    \caption{Traffic congestion }
    \label{fig:congestion}
\end{figure}

Table \ref{table:performanceXGB} reports the performance of the XGBoost-based prediction model, where the achieved \textit{mean squared error} (MSE) and \textit{mean absolute error} (MAE) vary in the ranges: [0.0033-0.0080] and [0.05-0.11], respectively (Appendix provides more prediction results using Google Cluster Dataset to further validate the XGBoost's performance). The prediction error increases and training time elapsed (TTE) decreases with the size of the prediction interval because of decreased number of data samples.  Further, the achieved traffic estimation errors: MSE and MAE  are compared against different versions of FMLA \cite{zhang2022efficient} including standard network (FMLA-s), wide network (FMLA-w), deep network (FMLA-d), and Optimal case (prediction error is zero) as depicted in Fig. \ref{fig:my_label}. MAE and MSE obtained for Milan (viz., $Mi^{MAE}$ and $Mi^{MSE}$) are lesser for both the proposed and FMLA, however, significant reduction in error of 90.8\% is observed for Trentino in case of TM-CRN over FMLA-d because of intuitive pattern learning and optimization of extreme gradient boosting approach. 
\begin{table}[!htbp]
	\centering
	\caption[Table caption text] {XGBoost-based Prediction performance metrics}  
	\label{table:performanceXGB}
	\small
	\resizebox{4cm}{!}{
				\begin{tabular}{l c c c}
			\hline
			PWS (min)& MSE & MAE & TTE \\ \hline 		
			5&0.0038&0.0547&28.89 \\ 
			10&0.0047&0.0857&24.09 \\ 
			30&0.0056&0.0860.2&21.64\\ 
			60&30.0078&0.1055&19.95\\ \hline
			\end{tabular}}\\ {\footnotesize{PWS: Prediction window size, TTE: training time elapsed (msec)}}
\end{table}
Fig. \ref{fig:congestion} compares TM-CRN with OSC-MC \cite{saxena2021osc} and Optimal case for occurrence of C-RAN traffic congestion using Youtube mobile streaming dataset.  TM-CRN reduces the congestion up to 10.15\%  over OSC-MC due to engagement of improved prediction capability and effective  distribution and management of the live traffic into distinct traffic states. TM-CRN is closer to Optimal for $Tr^{{25\%}}_{{25\%}}$, $Tr^{{50\%}}_{{50\%}}$ and show 2\%-9\% more congestion as compared with remaining cases of traffic thresholds.

\begin{figure}[!htbp]
\centering
	    \begin{tikzpicture}
\begin{axis}[
 width=.38\textwidth,
 height=0.24\textwidth,
    ymin=5,
    ymax=30,
    symbolic x coords={ 5\%, 25\%, 50\%,  75\%, 90\%},
    xtick=data,
    ybar,
    ymajorgrids=true,
    xmajorgrids=true,
    grid style=dashdotted,
    bar width=3.5pt,
    enlarge x limits=0.16,
    legend style={at={(1.2,0.98)},
    anchor=north,legend columns=1},
     xlabel={{\textit{CRAN-Traffic} ${BW^{thr}}$}},
    ylabel={$BW$ \textit{Utilization} (\%)}]
   \addplot+[ybar,fill=orange!10!,thin, mark options={fill=white},pattern color=orange!40!, draw=orange,thick, postaction={
        pattern=crosshatch, thin}, thin] plot coordinates{ ({5\%}, 15.24) ({25\%}, 22.50) ({50\%}, 28.059) ({75\%}, 29.7) ({90\%}, 24.57)}; \addlegendentry{\small{Optimal}} 
   
    \addplot+[ybar,fill=cyan!10!,thin, mark options={fill=white},pattern color=blue!40!, draw=cyan,thick, postaction={
        pattern=crosshatch, thin}, thin] plot coordinates{ ({5\%}, 13.24) ({25\%}, 19.50) ({50\%}, 26.159) ({75\%}, 27.07) ({90\%}, 22.57)}; \addlegendentry{\small{TM-CRN}} 
 \addplot+[ybar,fill=black!10!,thin, mark options={fill=black!10!}, postaction={
        pattern=crosshatch, thin},pattern color=black!50!, draw=black!60!, thin] plot coordinates{ ({5\%}, 16.06) ({25\%}, 14.06) ({50\%}, 13.06) ({75\%}, 12.00) ({90\%}, 10.06) }; \addlegendentry{\small{JUFO \cite{kim2019task}}}         

 \addplot+[ybar,fill=blue!6!,thin, mark options={fill=black!10!}, postaction={
        pattern=crosshatch, thin},pattern color=blue!50!, draw=blue!70!, thin] plot coordinates{ ({5\%}, 11.765) ({25\%}, 11.765) ({50\%}, 11.765) ({75\%}, 11.765) ({90\%}, 11.765)  }; \addlegendentry{\small{Rand-Fit}}     
   \addplot+[ybar,fill=teal!20!,thin, mark options={fill=cyan}, pattern color=teal!70!, draw=teal!100!, thick,postaction={
        pattern=crosshatch, thin}, thin] plot coordinates{ ({5\%}, 11.075) ({25\%}, 11.075) ({50\%}, 11.075) ({75\%}, 11.075) ({90\%}, 11.075) }; \addlegendentry{\small{First-Fit}}
  
\end{axis}
\end{tikzpicture}
	    \caption{Network bandwidth utilization }
	    \label{fig:bw}
	\end{figure}
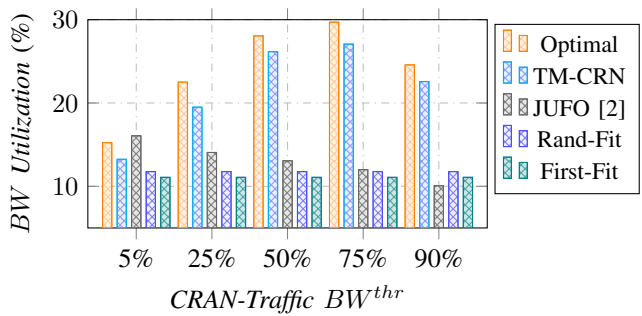

Fig. \ref{fig:bw} compares the network bandwidth utilization percent of TM-CRN with Optimal case, JUFO \cite{kim2019task} and baseline methods: random-fit and first-fit over varying  thresholds for both $BW^{thr}$ and $Et^{thr}$ viz., 5\%, 25\%, 75\%, and 90\%  during experiments. The bandwidth utilization has improved by 17.07\%, 17.31\%, and 18\% over JUFO, random-fit, and first-fit, respectively and reduced by 2.7\% over Optimal case for $BW^{thr}$ = $Et^{thr}$ = 75\% because of the division of the jobs according to their bandwidth and execution time requirement into distinct traffic states followed by the privileged management accordingly that accelerates the optimization of C-RAN jobs allocation and execution.

\begin{figure}[!htbp]
    \centering
    \begin{tikzpicture}[node distance = 1cm,auto,scale=.90, transform shape]
\begin{axis}[
 width=.40\textwidth,
 height=0.26\textwidth,
    ymin=100000,
    ymax=255000,
    symbolic x coords={ 5\%, 25\%, 50\%,  75\%, 90\%},
    xtick=data,
    ybar,
    ymajorgrids=true,
    xmajorgrids=true,
    grid style=dashdotted,
    bar width=3.5pt,
    enlarge x limits=0.16,
    legend style={at={(1.18,0.98)},
    anchor=north,legend columns=1},
     xlabel={{\textit{CRAN-Traffic} ${Et^{thr}}$}},
    ylabel={$Et$ \textit{consumption} (sec)}]
    \addplot+[ybar,fill=orange!10!,thin, mark options={fill=white},pattern color=orange!40!, draw=orange,thick, postaction={
        pattern=crosshatch, thin}, thin] plot coordinates{ ({5\%}, 114045) ({25\%}, 114277) ({50\%}, 110623) ({75\%}, 114815) ({90\%}, 119116)}; \addlegendentry{\small{Optimal}} 
    \addplot+[ybar,fill=cyan!10!,thin, mark options={fill=white},pattern color=blue!40!, draw=cyan,thick, postaction={
        pattern=crosshatch, thin}, thin] plot coordinates{ ({5\%}, 234045) ({25\%}, 187277) ({50\%}, 200623) ({75\%}, 194815) ({90\%}, 179116)}; \addlegendentry{\small{TM-CRN}} 
 \addplot+[ybar,fill=black!10!,thin, mark options={fill=black!10!}, postaction={
        pattern=crosshatch, thin},pattern color=black!50!, draw=black!60!, thin] plot coordinates{ ({5\%}, 244763) ({25\%},242763) ({50\%}, 234763) ({75\%}, 222045) ({90\%}, 239045) }; \addlegendentry{\small{JUFO \cite{kim2019task}}}         

 \addplot+[ybar,fill=blue!6!,thin, mark options={fill=black!10!}, postaction={
        pattern=crosshatch, thin},pattern color=blue!50!, draw=blue!70!, thin] plot coordinates{ ({5\%}, 245050) ({25\%}, 245050) ({50\%}, 245050) ({75\%}, 245050) ({90\%}, 245050)  }; \addlegendentry{\small{Rand-Fit}}     
   \addplot+[ybar,fill=teal!20!,thin, mark options={fill=cyan}, pattern color=teal!70!, draw=teal!100!, thick,postaction={
        pattern=crosshatch, thin}, thin] plot coordinates{ ({5\%}, 245765) ({25\%}, 245765) ({50\%}, 245765) ({75\%}, 245765) ({90\%}, 245765) }; \addlegendentry{\small{First-Fit}}
 \end{axis}
\end{tikzpicture}

    \caption{Service execution time}
    \label{fig:Et}
\end{figure}
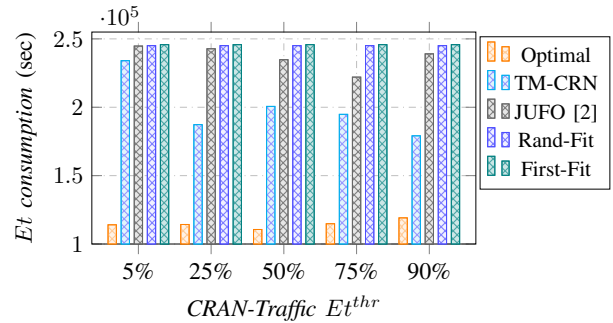

The comparison of execution time consumed is shown in Fig. \ref{fig:Et}, the proposed approach of C-RAN traffic management consumes 19.33\%, 26.90\%, and  27.12\% lesser time over JUFO, random-fit, and first-fit, respectively. Accordingly, the job execution latency for TM-CRN management is reduced by 23.80 \%, 18.37\%, 20.73\%, and 27.12\% for the $BW^{thr}$ and $Et^{thr}$ values: 25\%, 50\%, 75\%, and 90\%, respectively over default FCFS scheduling-based traffic management.  On the other hand, the job latency has been reduced up to 1.22 \%, 10.68\%, and 2.74\% for JUFO. However, the time elapsed for TM-CRN over Optimal case is higher up to 33.5\%-51.27\% because of the intended traffic prediction and analysis in case of TM-CRN which is avoided for Optimal case. The significant time reduction is achieved for the proposed approach due to the traffic classification into distinct  states, and incorporation and exploitation of parallelism by executing hogs having higher bandwidth and lesser execution time requirement in-between  stragglers (slow executing jobs).                                                                                                                                                                                                                                                                                                                                                                  
	
\section{Conclusion}
A novel TM-CRN model is proposed  to facilitate effective RAN traffic management with minimum latency period and maximum bandwidth utilization. The probable resource demand and  traffic congestion  is detected using XGBoost predictor. Traffic management unit classifies that traffic into distinct  states to  utilize available bandwidth productively for parallel execution of RAN traffic confined to stragglers and network hogs.   The performance evaluation and comparison confirmed that TM-CRN model is more admissible as compared with state-of-the-art approaches.

\bibliographystyle{IEEEtran} 
\bibliography{bibfile}

\begin{thebibliography}{1}
\providecommand{\url}[1]{#1}
\csname url@samestyle\endcsname
\providecommand{\newblock}{\relax}
\providecommand{\bibinfo}[2]{#2}
\providecommand{\BIBentrySTDinterwordspacing}{\spaceskip=0pt\relax}
\providecommand{\BIBentryALTinterwordstretchfactor}{4}
\providecommand{\BIBentryALTinterwordspacing}{\spaceskip=\fontdimen2\font plus
\BIBentryALTinterwordstretchfactor\fontdimen3\font minus
  \fontdimen4\font\relax}
\providecommand{\BIBforeignlanguage}[2]{{%
\expandafter\ifx\csname l@#1\endcsname\relax
\typeout{** WARNING: IEEEtran.bst: No hyphenation pattern has been}%
\typeout{** loaded for the language `#1'. Using the pattern for}%
\typeout{** the default language instead.}%
\else
\language=\csname l@#1\endcsname
\fi
#2}}
\providecommand{\BIBdecl}{\relax}
\BIBdecl

\bibitem{zhao2020federated}
Z.~Zhao, C.~Feng, H.~H. Yang, and X.~Luo, ``Federated-learning-enabled
  intelligent fog radio access networks: Fundamental theory, key techniques,
  and future trends,'' \emph{IEEE wireless comm.}, vol.~27, no.~2, pp. 22--28,
  2020.

\bibitem{kim2019task}
J.~Kim, T.~Ha, W.~Yoo, and J.-M. Chung, ``Task popularity-based energy
  minimized computation offloading for fog computing wireless networks,''
  \emph{IEEE Wireless Comm. Letters}, vol.~8, no.~4, pp. 1200--1203, 2019.

\bibitem{habibi2019comprehensive}
M.~A. Habibi, M.~Nasimi, B.~Han, and H.~D. Schotten, ``A comprehensive survey
  of ran architectures toward 5g mobile communication system,'' \emph{IEEE
  Access}, vol.~7, pp. 70\,371--70\,421, 2019.

\bibitem{chen2016xgboost}
T.~Chen and C.~Guestrin, ``Xgboost: A scalable tree boosting system,'' in
  \emph{Proceedings of the 22nd acm sigkdd international conference on
  knowledge discovery and data mining}, 2016, pp. 785--794.

\bibitem{saxena2022fault}
D.~Saxena, I.~Gupta, A.~K. Singh, and C.-N. Lee, ``A fault tolerant elastic
  resource management framework towards high availability of cloud services,''
  \emph{IEEE Transactions on Network and Service Management}, 2022.

\bibitem{barlacchi2015multi}
G.~Barlacchi, M.~De~Nadai, R.~Larcher, A.~Casella, C.~Chitic, G.~Torrisi,
  F.~Antonelli, A.~Vespignani, A.~Pentland, and B.~Lepri, ``A multi-source
  dataset of urban life in the city of milan and the province of trentino,''
  \emph{Scientific data}, vol.~2, no.~1, pp. 1--15, 2015.

\bibitem{lohYoutube2022}
F.~Loh, F.~Wamser, F.~Poignée, S.~Geißler, and T.~Hoßfeld, ``Youtube dataset
  on mobile streaming for internet traffic modeling, network management, and
  streaming analysis. figshare. dataset,''
  \emph{https://doi.org/10.6084/m9.figshare.19096823.v2}, 2022.

\bibitem{zhang2022efficient}
L.~Zhang, C.~Zhang, and B.~Shihada, ``Efficient wireless traffic prediction at
  the edge: A federated meta-learning approach,'' \emph{IEEE Comm. Letters},
  2022.

\bibitem{saxena2021osc}
D.~Saxena and A.~K. Singh, ``{OSC-MC}: Online secure communication model
  for cloud environment,'' \emph{IEEE Comm. Letters}, vol.~25, no.~9, pp.
  2844--2848, 2021.

\end{thebibliography}

\end{document}